\documentclass[12pt]{article}
\usepackage{mathrsfs}
\usepackage{amsmath,amssymb,array}
\usepackage{graphicx}
\usepackage{bbm}
\usepackage{multicol}
\newcommand{\cN}{{\cal N}}
\numberwithin{equation}{section}
\def\p{\partial}

\begin{document}

\begin{titlepage}
\renewcommand{\thefootnote}{\fnsymbol{footnote}}

\begin{center}
{\LARGE \bf Note on the Mesons Mass Spectrum in a Soft-Wall AdS/QCD Model}

\vspace{1.0cm}

{Keita Kaniba Mady$^{\dag}$}\footnote{E-mail: madyfalaye@gmail.com},
{Dicko Younouss Ham\`{e}ye$^{\dag,\ddag}$}
\\

\vspace{.5cm}
{\it\small {$^{\dag}$ D\'{e}partement de Physique, Facult\'{e} des Sciences et Techniques, Universit\'{e} des Sciences, des Techniques et des Technologies de Bamako, Mali,}\\ $^{\ddag}$ Ecole Superieure de Technologie et de Management, ESTM, Mali}\\

\vspace{.3cm}
\today
\end{center}\vspace{1.5cm}

\centerline{\textbf{Abstract}}\vspace{0.5cm}

The primary goal of this paper is to analyze the mass spectrum of the vector ($\rho$) mesons by using the soft-wall AdS/QCD device.
The theory is characterized by the form $\Phi=\lambda^2z^2+\lambda z$ of the dilaton field, which satisfies the constraint for having the correct Regge behavior and avoids the presence of the spurious massless scalar mode in the spectrum. The spectrum we obtained is in good agreement with experimental data.
\end{titlepage}
\setcounter{footnote}{0}

%%%%%%%%%%%%%%%%%%%%%%%%%%%%%%%%%%%%%%%%%%%%%%%%%%%%%%%%%%%%%%%%%%%%%%%%%%
%%%%%%%%%%%%%%%%%%%%%%%%%%%%%%%%%%%%%%%%%%%%%%%%%%%%%%%%%%%%%%%%%%%%%%%%%%
%%%%%%%%%%%%%%%%%%%%%%%%%%%%%%%%%%%%%%%%%%%%%%%%%%%%%%%%%%%%%%%%%%%%%%%%%%

\section{Introduction}

The elementary fermions that make up the proton and the neutron are called quarks. These fundamental particles are strongly interacting and have been successfully described by a non-Abelian color gauge theory called quantum chromodynamics, or QCD for short. In QCD the gauge group is the color $SU(3)_{c}$, and quarks $\psi$ and the quanta of the gauge fields (gluons) $A^{a}_{\mu}$ belong to its fundamental and adjoint representation, respectively.
A remarkable feature of QCD is the self-interactions among the $SU(3)_{c}$ gauge fields. It is a well-known fact that this self-interactions of the gauge fields is the main source of its asymptotic freedom behavior.  In the high energy regime, the quarks interact weakly, and one can  safely use the standard techniques of perturbation theory to discuss QCD at this regime. On the other hand, however, at the low energy regime, the quarks  are bounded inside the mesons and baryons as quark-antiquark or three quarks states with net color charge being zero. These bounded state of quarks are usually called hadrons. The quarks inside the hadrons are strongly interacting system. Consequently, there is a lack of analytically method to discuss from first principles the low energy properties of QCD, since the standard techniques of perturbation theory only apply to a weakly interacting system.

The solvability of QCD at low energy regions is plagued by a lack of expansion parameter. To overcome this issue, 't Hooft \cite{tH} suggested that one should learn properties about QCD from $SU(N)$ gauge theory in the limit of $N\mapsto \infty$. Since at this limit, $1/N$ can be used as an expansion parameter and a perturbative  study of QCD can be performed.  In addition, studies of this large N expansion suggested that there must be a correspondence between strongly coupled conformal field theories and gauge theories living in a higher dimensional AdS background geometry. This duality, original propounded by Maldacina \cite{M}, preaches the equivalence between the low energy approximation of type $IIB$ string theory on $AdS_{5}\bigotimes S^{5}$ and  $\cN=4$ U(N) SYM theory for large $N$ in four dimensions. This Anti-de Sitter/conformal field theory (AdS/CFT) correspondence\cite{GKP, OGO, W}, has been widely studied at length in the literature and it advocates the correspondence between a strongly conformal theory in $d$ dimensional spacetime and a gauge theory formulated in $AdS_{d+1}$ background. Although QCD is not a conformal theory, we believe that we can still apply the AdS/CFT recipe to  it due to its conformal behavior at high energy regime. This phenomenological cousin of AdS/CFT has been termed AdS/QCD.

In many respects, the AdS/QCD is an ideal pedagogical device for learning about the low energy properties of QCD (the mass spectrum, the form factors, the correlator functions, the coupling constant and the decay constant of the bounded state of quarks). The fundamental lesson preached by the AdS/QCD duality is the equivalence between theory of gravity living in a five-dimensional AdS geometry and low energy QCD at the boundary of this geometrical background. This is mainly due to two basic tenets: On one hand the boundary of the five-dimensional AdS model is a four-dimensional spacetime that looks like flat spacetime with three spatial directions and one time direction; and on the other hand Yang-Mills theories at the boundary of $AdS_{5}$ is equivalent to gravitational physics in the $AdS_{5}$ geometry. There are two approaches to study QCD through AdS/QCD: the hard-wall model \cite{dTB, EKSS, DP1} and the soft-wall model \cite{KKSS}.
In this paper, we limit ourselves to the soft-wall model. Specifically, we analyze the mass spectrum of the vector ($\rho$) mesons. The virtue of soft-wall AdS/QCD model is that it achieves linear confinement \cite{KKSS, Z, SWY} and chiral symmetry breaking \cite{GKK, DP1, VS2} in a very simple way.

Over the recent years, by using the soft wall AdS/QCD device, several researchers have theoretically investigated \cite{KKSS, Z, VS1, VS2, GKK, SWY, WF} the mass spectra of the resonance messons sector of the low energy QCD. Nevertheless, the discrepancy between the theoretical value and the experimental value of some of the masses is still very large.

The potential of the Schr$\ddot{o}$dinger like equation that determines the mass spectrum of the $\rho$ mesons is the simplest of all the messons sector, since it only depends on the dilaton field $\Phi(z)$ and the conformal factor $a(z)$. So we believe that if on arrives to reduce the discrepancy for the $\rho$ messon mass, the others sectors should be reduce by a suitable parameterizations of the vev of the scalar field $X(z)$. Therefore,  It would seem reasonable to redo the analysis of the $\rho$ meson sector again.

The rest of this paper is organized as follows: The next section opens with a review of the soft-wall AdS/QCD. In section 3, we present our model and the mass spectrum it gives. The final section is devoted to the conclusion.

\section{The Soft-Wall AdS/QCD model}

The soft-wall AdS/QCD is the bottom-up approach to QCD in AdS/QCD correspondence. That is, one starts from QCD and constructs its  gravity dual theory.  For an understanding of how it works see ref(\cite{JP}).

 For an illustration of this bottom-up approach, let us consider the Dirac Lagrangian that governs the motion of quarks, that is,
\begin{equation}\label{dirac}
    L_{Dirac}=\bar{\psi}(i\gamma^{\mu}\partial_{\mu}-m)\psi.
\end{equation}
In term of the Chiral components, \begin{eqnarray}
\psi=\begin{pmatrix}\,\, \psi_{R}  \,\,\,\, \\[0.2cm]
                           \,\, \psi_{L}  \,\,\,\, \end{pmatrix} \,\,,\nonumber
\end{eqnarray} equation (\ref{dirac}) can be rewritten as:
\begin{equation}\label{chiral}
   L_{Dirac}=\bar{\psi}_{L}i\gamma^{\mu}\partial_{\mu}\psi_{L}+\bar{\psi}_{R}i\gamma^{\mu}\partial_{\mu}\psi_{R}-m(\bar{\psi}_{R}\psi_{L}+\bar{\psi}_{L}\psi_{R}).
\end{equation}

It can be demonstrated that this lagrangian has a global $SU(N_f)_L \times SU(N_f)_R$ symmetry in the massless limit (m=0)\footnote{$N_{f}$ is the number of quarks flavors}. This Chiral symmetry is explicitly broken by the mass term. Additionally, there is another source, the so-called quarks condensate $\langle\bar{\psi}(x)\psi(x)\rangle$, that spontaneously breaks the Chiral symmetry. Furthermore, the spectra of mesons in low energy QCD are linear and the quark are confined inside the hadrons. This linear confinement feature has to be realized in any realistic model of low energy QCD.

The building of the holographic (gravity) dual of QCD starts with these basic ingredients of low energy QCD. According to the dictionary, each global symmetry of QCD becomes a local symmetry in the gravity side. Thus, in the gravity side, there is one left-handed gauge vector field (A$_{L}$) for the global $SU(N_{f})_{L}$ symmetry of QCD and one right-handed gauge vector field (A$_{R}$) for the global $SU(N_{f})_{R}$ symmetry of QCD. The spontaneously breaking of the chiral symmetry is achieved by introducing a bifundamental scalar $X(z)$ field, which belong to the adjoint representation of the 5D gauge group $SU(N_f)_L \times SU(N_f)_R$. To satisfy the linear confinement feature of QCD, one simply has to turn on a holographic coordinate dependent dilaton field $\Phi(z)$ with the limit $\Phi(z\rightarrow\infty)\sim \,\lambda^2 z^2\, $ \cite{KKSS}. Finally and more importantly, the background in which these bulk fields live is assume to be a slice of AdS spacetime. The exact mapping between the fields in the two side is depicted in table $1$. The holographic coordinate $z$, which corresponds to the energy scale in the four-dimensional theory, is defined within the range $0<z<\infty$.

\begin{table}
\begin{center}
\caption{The mapping between the fields in the two side and their dimensions.}
\begin{tabular}{ c c c | c c c | c c c | c c c }
  \hline
  \hline
  & 4D : \textit{O}(x) & & & 5D : $\Phi^{bulk}(x,z)$ & & & $\Delta$ & & & $m_{5}^{2}$ & \\
  \hline
  & $\overline{\psi}_{L} \gamma^{\mu} t^{a} \psi_{L}$ & & & $A_{L \mu}^{a}$ & & & 3 & & & 0 & \\
  & $\overline{\psi}_{R} \gamma^{\mu} t^{a} \psi_{R}$ & & & $A_{R \mu}^{a}$ & & & 3 & \\
  & $\overline{\psi}_{R}^{\alpha} \psi_{L}^{\beta}$ & & & $\frac{1}{z} X^{\alpha\beta}$& & & $3$ & & & $-3$ & \\
  \hline
  \hline
\end{tabular}
\end{center}
\end{table}

The 5D non-renormalizable bulk action\footnote{ We limit ourselves to the quadratic parts of the fields}, writing out of these fields, that describes the meson sector reads
\begin{eqnarray}
S_M=\int d^4x\,dz\, \sqrt{G}\,e^{-\Phi}\, \mathrm{Tr}\left\{-\frac{1}{4g_5^2}(\,F_L^2+F_R^2)+
    |DX|^2-m_X^2|X|^2\,\right\}\, \label{SM},
\end{eqnarray}
where $F_L$ and $F_R$ are the nonabelian field strength formed from the gauge potentials (A$_{L}$) and (A$_{R}$) respectively and are defined by
\begin{equation}
 F_{L,R}^{MN} = \partial^{M} A_{L,R}^{N} - \partial^{N} A_{L,R}^{M} - i [A_{L,R}^{M},A_{L,R}^{N}]. \nonumber
\end{equation}
The symbol $D$ is the Yang-Mills covariant derivative containing the gauge fields (A$_{L}$, A$_{R}$). That is, $D_M{X}=\p_M{X}-iA_{LM}{X}+i{X}A_{RM}$ and
the AdS/QCD dictionary fixes the gauge coupling $g_5$ with QCD to be $g_5^2=12\pi^2/N_c=4\pi^2$ (for $N_c=3$ as in QCD).
The metric is an AdS geometrical background
\begin{eqnarray}
ds^2=\,G_{MN}\,dx^M dx^N=a(z)\,(\,\eta_{\mu\nu}dx^{\mu}dx^{\nu}-dz^2),\label{metric},
\end{eqnarray} where the 4D Minkowski metric is given by $(\eta_{\mu\nu})=diag(1,-1,-1,-1)$, and $a(z)$ is the conformal factor or the warped factor.

The vacuum expectation value(vev) of the scalar field ($\langle{X}\rangle=\frac{1}{2}\,v(z)$) is assumed to have a $z$-dependent with the limit
\begin{eqnarray}
v(z\rightarrow0)\sim \,\alpha z+\beta z^3\,. \label{vUV}\nonumber
\end{eqnarray}
From (\ref{SM}), the equation of motion (EOM)  of the vev $v(z)$, in the axial gauge to be defined below, reads
\begin{eqnarray}
\p_z(\,a^3 e^{-\Phi}\p_z v)+3a^5 e^{-\Phi}v=0\,. \label{EOMv}
\end{eqnarray}

The hadrons are defined as the renormalizable modes of the 5D gauge fields. For instance,
the vector mesons (V) and the axial-vector mesons (A) are respectively define by:
\begin{eqnarray}
  V=(A_{L}+A_{R})/2\,,\qquad  A=(A_{L}-A_{R})/2
 \nonumber
\end{eqnarray}

An analysis of gauge theory is usually simplified by choosing a gauge fixing term. Here, we choose the most commonly used gauge fixing term, that is, the axial gauge $V_z=0$.

It is a well-known fact that the Kalusa-Klein dimensional reduction of a 5D vector field gives rise to an infinite tower of 4D massive vector fields called the (KK) modes. This KK decomposition is usually done by breaking the field of interest into an infinite tower of 4D component satisfying the Proca's equation and purely extra dimension dependent parts. For example, one can decompose the vector meson field (V) as $V_\mu(x,z)=\sum_{n}\,\rho_\mu^{(n)}(x)h_V^{(n)}(z)$. The z-dependent parts satisfy the constraint \cite{KM}
\begin{equation}\label{eigen}
    -\p_5(ae^{-\Phi}\p_5h_V^{(n)})=ae^{-\Phi}M_V^{(n)2}h _V^{(n)},
\end{equation} where $M_V^{(n)2}$ are the masses of the vector fields $\rho_\mu^{(n)}(x)$.

The infinite tower of 4D massive vector fields $\rho_\mu^{(n)}(x)$ resulting from  the Kalusa-Klein decomposition of the vector field $V$ are assumed to be the vector $\rho$ mesons of low energy QCD and the equation (\ref{eigen}) determines its mass spectrum.

%\begin{eqnarray}
%V_\mu(x,z)=\sum_{n=0}^{\infty}\,\rho_\mu^{(n)}(x)\,f_V^{(n)}(z)\,,\label{KKr}
%\end{eqnarray}
%%%%%%%%%%%%%%%%%%%%%%%%%%%%%%%%%%%%%%%%%%%%%%%%%%%%%%%%%%%%%%%%%%%%%%%%%%%%%%%%%
%\begin{eqnarray}
%&&-\,\p_5(ae^{-\Phi}\p_5f_V^{(n)})=ae^{-\Phi}M_V^{(n)2}f_V^{(n)}\,,\\%[0.2cm]
%&&\hspace{0.3cm} f_V^{(n)}|_{z\rightarrow0}=0\,,\quad\quad f_V^{(n)}|_{z\rightarrow\infty}=0\,.\label{Mr}
%\end{eqnarray}

Equation (\ref{eigen}) can be deformed, by simply setting $h_V^{(n)}=e^{[\Phi(z)-\log{a(z)}]/2}\chi_V^{(n)}$, to a Schr\"{o}dinger form
\begin{equation}\label{sch}
  -\chi_V^{(n)\prime\prime}+V_V\chi_V^{(n)}=M_V^{(n)2}\chi_V^{(n)}
\end{equation}
 with the potential
\begin{eqnarray}
V_V=\frac{1}{4}[\Phi(z)-\log{a(z)}]'^{\,2}-\frac{1}{2}[\Phi(z)-\log{a(z)}]''\,,\label{VV}
\end{eqnarray} where ($'$) denotes derivation with respect to z.

\section{The Model and its Parameters}

The analysis we intend to do is based on a modified version of the soft-wall model advocated by Erlich et al.\cite{EKSS} and further scrutinize in \cite{GKK, Z, VS1, KKSS2}. We assume that the bulk fields are propagating on the boundary of a slice 5D AdS background geometry, where the metric in the Poincar\'{e} coordinates is given by equation (\ref{metric}).

With the constraint on the dilaton field in our mind, and the fact that we should recovered the AdS geometry in the UV regime, that is, $a(z\rightarrow0)\sim \,L/z\, $, we define our model in table 2.
\begin{table}
\begin{center}
\caption{The form of the dilaton field and the conformal factor used in this model.}
\begin{tabular}{|cc|cc|cc|cc|}
                   \hline
                   A & &  $\Phi_{A}(z)= \lambda^2 z^2+\lambda z$ & & $a_{A}(z)=1/z$ & \\
                   B &&  $\Phi_{B}(z)= \lambda^2 z^2+\lambda z$ && $a_{B}(z)=(m+n z^2)/z$ &\\
                   C &&  $\Phi_{C}(z)= \lambda^2 z^2+\lambda z$ && $a_{C}(z)=\sqrt{\alpha^2+d z^2}/z$ & \\
  \hline
\end{tabular}
\end{center}
\end{table}
Comparatively, the form of our dilation field $\Phi(z)= \lambda^2 z^2+\lambda z$ is different to that of \cite{HG} in two points. First thing to notice in our model is that it simply needs one parameter in contrast to two in \cite{HG}. Finally and more importantly, the sign of the z-dependent tern is positive in our model. This is very important, since it has been shown in \cite{KKSS2} that,in order to avoid the presence of a spurious massless state in the vector sector, the sign of the exponential profile defining the wall should be positive.

Using (\ref{VV}) and the parametrization given in table 2, One can easily evaluated the potentials of the Schr\"{o}dinger-like problem, and the results are:
\begin{flushleft}
 \begin{equation}\label{Model A}
    V_{VA}(x)=\lambda^4 x^2+\lambda^3 x+\frac{\lambda^2}{4}+\frac{\lambda^2}{2 x}+\frac{3}{4 x^2},\nonumber
 \end{equation}
\end{flushleft}
\begin{multline}\label{Model B}
    V_{VB}(x)=\lambda^4 x^2+\lambda^3 x-\lambda^2\frac{n x^2-m}{n x^2+m}-
     \frac{3 \lambda^2}{4}-\frac{n \lambda x^2-m \lambda-n x}{2 n x^3+2 m x}\\+
     \frac{m}{2 x^2(n x^2+m)}-\frac{3 n^2 x^2}{4 (n x^2+m)^2}+\frac{mn}{2 (n x^2+m)^2}+\frac{m^2}{4 x^2(n x^2+m)^2},\nonumber
\end{multline}
\begin{equation}\label{Model c}
   V_{VC}(x)=\lambda^4 x^2+\lambda^3 x+\frac{\lambda^2}{4}+\frac{\lambda^2}{2 x}+\frac{3}{4 x^2}-d x \lambda\frac{(2 d x \lambda+1)}{2 \alpha^2+2 d x^2}-3 \frac{d^2 x^2}{4(\alpha^2+d x^2)^2}.
\end{equation}
 To get the mass spectrum from the Schr\"{o}dinger equation (\ref{sch}) with the potentials given in (\ref{Model c}), we used the shooting method with the boundaries $\chi_n(z\to 0) =0$, $\partial_z
\chi_n(z\to \infty) =0$; the obtained mass spectra are presented in table 3.

The values of the five parameters that fit the experimental data are:
\begin{eqnarray}
&\lambda=401.57\,\mathrm{MeV}\,,\quad \alpha=0.7\,\mathrm{MeV}\,;& \nonumber \\
&m=1.1\,\mathrm{MeV}\,,\quad n=6000\,\mathrm{MeV}\,,\quad d=5900\,\mathrm{MeV}\,.& \label{para_m}
\end{eqnarray}

\begin{table}
\begin{tabular}{|c|c|c|c|c|c|c|c|c|}
\hline
$\rho$  mass (MeV)                  &  0  &  1   &  2   &  3   &  4   &  5   &  6   \\
\hline
Model A  &983.6 & 1297 & 1544  & 1755 & 1941.6 & 2140  & 2266.8 \\
\hline
Model B  &970 & 1280& 1525 & 1734 & 1920 & 2113 & 2244 \\
\hline
 Model C &970 & 1281& 1527 &1737 & 1924 & 2093 & 2287\\
\hline
experimental   &775.5$\pm$ 1& 1282$\pm$ 37 & 1465$\pm$ 25 & 1720$\pm$ 20 & 1909$\pm$30 & 2149$\pm$ 17 & 2265$\pm$ 40 \\
\hline
error (Model A) &26.8\%& 1.2\% &5.4\% & 2\% & 1.7\% & 0.4\% & 0.1\% \\
\hline
error (Model B) &25\%& 0.2\% &4.1\% & 0.8\% & 0.6\% & 1.7\% & 0.9\% \\
\hline
error  (Model C) &25\%& 0.1\% &4.2\% & 1\% & 0.8\% & 2.6\% & 1\%\\
\hline
\end{tabular}
\caption{\small{The theoretical and experimental values of the masses of the vector $\rho$ meson in the three cases.}}\label{rho}
\end{table}

\section{Conclusions}

In this paper, by using the device of the soft-wall AdS/QCD model, we analyzed the $\rho$ mass spectrum with the form of the dilation field $\Phi(z)= \lambda^2 z^2+\lambda z$. This form of the dilation field satisfies the constraint for having the correct Regge behavior as well as the constraint for the non-existence of the spurious massless scalar mode in the spectrum. Moreover, as it can be seen in table 3, there is an agreement between our model and the experimental data. Therefore, it seems important to extend our analysis by redoing the investigation of the other sectors using our dilaton field profile. Comparatively, in figure 1, we can see that the model $B$ as well as back-reacted geometry \cite{WF} (case C) approache very well to the experimental data. Consequently, in the soft wall AdS/QCD model, the back-reaction of the geometry has to be taken into account.

%\begin{acknowledgments}
\subsection*{Acknowledgments}
We wish to express our thanks to Alfredo Vega, Thomas Kelley, and Peng Zhang for useful correspondences. Keita is indebted to Wu Feng for his encouragement. He would also like to thank all the members of the department of physics of Nanchang University, and the teams of the Center for Relativistic Astrophysics and High Energy Physics of Nanchang University.

%\end{acknowledgments}

\newpage
\begin{figure}[t]
\begin{center}
\includegraphics[width=20cm,clip=true,keepaspectratio=true]{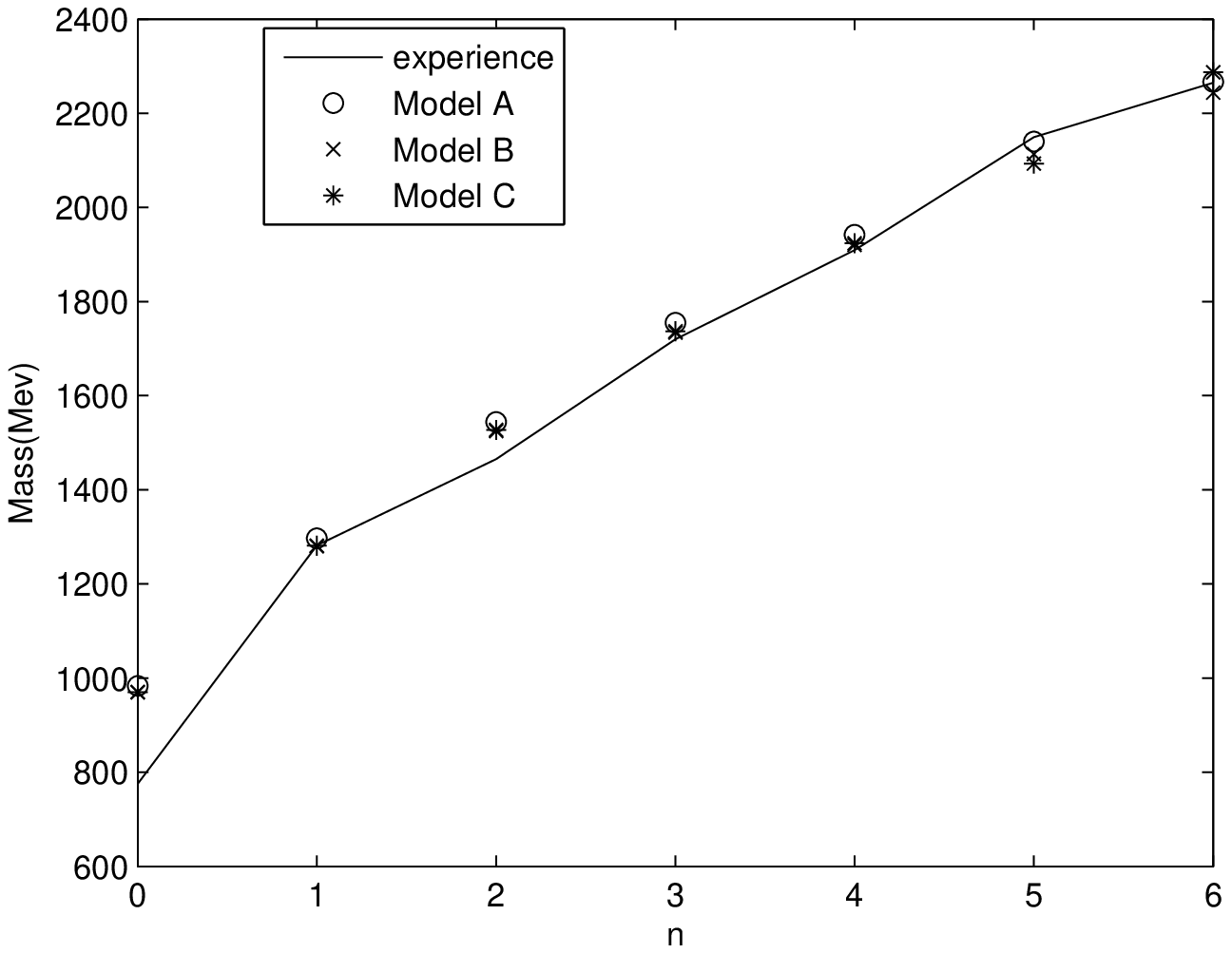}
\caption{\small The mass spectra of the different models compare to the experimental data.}
\end{center}\label{kk}
\end{figure}

\end{document}